\documentclass[reprint,aps,prl,amsmath,amssymb]{revtex4-1}
\usepackage{changes}

\usepackage[]{graphicx}
\usepackage{times}
\usepackage{bm}
\usepackage{color}

\newcommand{\comment}[1]{}

\begin{document}

\title{Optical fingerprint of dark 2p-states in transition metal dichalcogenides
}
\author{Gunnar Bergh\"auser$^{1}$, Andreas Knorr$^2$, and Ermin Malic$^1$} 
\email{ermin.malic@chalmers.de}
\affiliation{$1$ Department of Physics, Chalmers University of Technology, Gothenburg, Sweden, $2$ Institut f\"ur Theoretische Physik, Nichtlineare Optik und Quantenelektronik, Technische Universit\"at Berlin, Hardenbergstr. 36, 10623 Berlin, Germany}

\begin{abstract}
We present a microscopic study on higher excitonic states in transition metal dichalcogenides in the presence of disorder. We show that the geometric phase cancels the degeneration of 2s and 2p states and that a significant disorder-induced coupling of bright and dark states offers a strategy to circumvent optical selection rules. As a prove, we show a direct fingerprint of dark 2p states in absorption spectra of WS$_2$. The predicted softening of optical selection rules through exciton-disorder coupling is of general nature and therefore applicable to related two-dimensional semiconductors.\end{abstract}

\maketitle
Atomically thin transition metal dichalcogenides (TMDs) are in the focus of a rapidly growing scientific community due to their exceptional optical and electronic properties \cite{novoselov05, Butler2013}.
In contrast to two dimensional graphene, the inversion symmetry is broken, resulting in a band gap at the K and K' points \cite{HeinzPRL2010, Splendiani2010,Tonndorf2013}.
 In optical absorption spectra, the extraordinarily strong Coulomb interaction gives rise to the formation of excitonic and trionic states with binding energies in the range of 500 meV \cite{Ramasubramaniam2012,Berkelbach2013,Louie2013PRL,Steinhoff2014,Chernikov2014,Bergh2014b} and 30 meV \cite{Mak2013,Berkelbach2013,XiaoqinPRL2014}, respectively. Furthermore, disorder is known to play an important role in TMDs \cite{ImpuritiesOnTMDWallace2015} and is in the first place responsible for the broad inhomogeneous line widths in optical spectra that can exceed the homogeneous line width by one order of magnitude \cite{Moody2015,HuberKorn2015}.
So far, most theoretical studies of excitonic properties focus on ideal excitonic spectra \cite{Louie2013PRL,Steinhoff2014,Bergh2014b,MacDonald2015,Berkelbach2015}
however disorder effects have been found to play a significant role in optical experiments \cite{Moody2015}.\\
In this Letter, we use the Heisenberg's equation of motion formalism \cite{PQE} to investigate the impact of disorder on higher 2p like excitonic states. We address the question whether exciton-disorder coupling can circumvent optical selection rules and make these dark excitonic states optically active. TMDs are excellent candidates to investigate this generally interesting question, since they (i) show a significant inhomogeneous broadening of the absorption spectrum indicating a disorder landscape in the semiconductor, (ii) exhibit a strong Coulomb interaction resulting in experimentally accessible Rydberg-like series of higher excitonic states \cite{Chernikov2014}, and (iii) are characterized by a non-vanishing geometric phase, which cancels the degeneration of 2s and 2p excitonic states making both states detectable in optical spectra \cite{Barry_PRL_Imamoglu2015,Barry_PRL_Xiao2015,Barry_PRL_Louie2015,MacDonald2015,Berkelbach2015}. As a result, optical fingerprints for initially dark higher excitonic states are expected in optical spectra provided that the exciton-disorder coupling is efficient enough to soften the optical selection rules.

 \begin{figure}[t!]
 \begin{center}
\includegraphics[width=1.\linewidth]{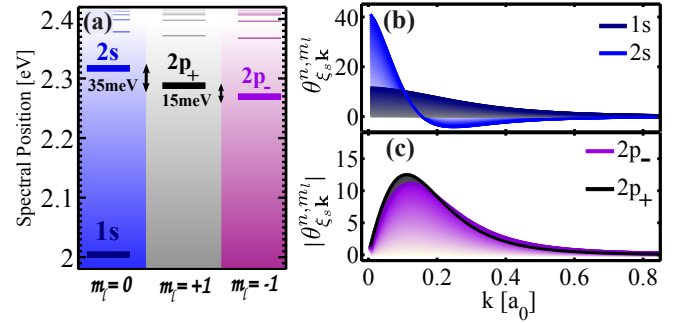}
 \end{center}
 \caption{ Excitonic states and wavefunctions in WS$_2$. (a) The extraordinarily strong Coulomb interaction leads to relatively large binding energies of higher excitonic states resulting in a Rydberg-like series. The non-vanishing geometric phase in TMDs lifts the energy degeneration of the azimuthal quantum number $m_l$. As a result, 2p states have a larger binding energy and are located below the 2s states in contrast to the hydrogen atom. Furthermore, we show momentum-dependent excitonic wave functions of the two energetically lowest (b) bright s states and (c) dark p states.}
\label{fig:1}
\end{figure}
To investigate the impact of exciton-disorder coupling on the absorption spectra\cite{BechstedtPRB1994,ZimmermannRunge1999}, we calculate the frequency-dependent absorbance $\alpha(\omega)$ that is proportional to the imaginary part of the linear response function $\chi(\omega)$ \cite{ErminsBuch}. 
The latter is determined by the microscopic polarization $p^{ij}=\langle a^{\dagger}_{j}a_{i} \rangle$ describing the optical transition between the states $j$ and $i$. 
Here, we have used the formalism of second quantization introducing the creation and annihilation operators $a^{\dagger}_{j}$ and $a_{i}$. 
The interacting states are described by the compound indices $i, j$ containing the electronic momentum $\mathbf k_{i}$, $\mathbf k_{j}$, the electronic band $\lambda=v,c$ (denoting the conduction and the valence band) as well as the spin valley index $\xi_{s}$. Note that we restrict our study to the energetically lowest A exciton and we do not consider intervalley interactions 
or doping effects but focus on a finite center-of-mass motion $\mathbf k_i\neq \mathbf k_j$. 

In a first step, we derive semiconductor Bloch equations for the entire class of TMDs within the Heisenberg picture based on the many-particle Hamilton operator consisting of the interaction-free carrier contribution $H_0$, the light-matter interaction $H_{l-m}$, the Coulomb interaction $H_{C}$, and the disorder interaction $H_{d}$. Then, we project the solution for the microscopic polarization $p^{vc}_{\xi_s\mathbf k_i \mathbf k_j}$ into an excitonic basis using the transformation
 $p^{vc}_{\xi_s\mathbf k}=\sum_{n,m_l}p^{n,m_l} \theta^{n_l,m_l}_{\xi_s\mathbf k}$
 with the excitonic polarization $p^{n,m_l}$ characterized by the excitonic quantum number $n$ and the angular momentum or azimuthal quantum number $m_l$. The appearing excitonic wave function $\theta^{n_l,m_l}_{\xi_s\mathbf k}$ can be determined with the Wannier equation \cite{PQE,Bergh2014b}. 
 Note that for the study of the Wannier equation we only need to consider microscopic polarizations containing the relative motion $\mathbf k=\alpha\mathbf k_i +\beta\mathbf k_j$ denoting direct optical transitions.

Before we further discuss the results of the Wannier equation (cf. Eq.\ref{eq:Wannier}), we first address 
the lifting of the degeneracy of 2s and 2p states, which is of crucial importance to be able to detect these states in optical spectra. 
The lifting has two origins, which are both accounted for in the present study: the momentum-dependent screening of the Keldish Coulomb potential and the geometrical phase that is present in two-dimensional systems with the optically active region located at the K points and not as in conventional semiconductors at the $\Gamma$ point.
While at the latter, the crystal symmetry effects cancel and one can find a perfect analogy between the excitonic and the two dimensional hydrogen problem, at the K points the excitonic states and the eigenfunctions are influenced by the lattice symmetry appearing as a geometric phase. It enters via the overlap integrals of the matrix elements \cite{Barry_PRL_Imamoglu2015,Barry_PRL_Xiao2015,Barry_PRL_Louie2015}.
In our approach, the phase is inherently included in the tight-binding wave functions \cite{Bergh2014b}
$\Psi^{\lambda,\xi_s}(\mathbf k,\mathbf r)=\frac{1}{\sqrt{N}}\sum_{\mathbf{R}_j, j}C_{j \mathbf k}^{\lambda\xi_s}\,e^{i\mathbf k\cdot \mathbf R_j}\phi_{j}^{\lambda,\xi_s}(\mathbf r -\mathbf R_{j}).$
Here, $\phi_{j}^{\lambda,\xi_s}(\mathbf r -\mathbf R_{j})$ is the orbital integral of the irreducible basis of the atom type j, which can be M denoting the transition metal or X denoting the chalcogen atom in TMD structures MX$_2$.
The wave function of TMDs explicitly includes the lattice symmetry via the tight-binding coefficients \cite{Bergh2014b} 
\begin{equation}
\label{eq:coef}
C_{M,\mathbf{k}}^{\lambda\xi_s}=C^{\lambda\xi_s}_{X,\mathbf{k}}g_{\mathbf{k}}^{{\lambda}\xi_s},\quad C_{X,\mathbf{k}}^{\lambda \xi_s}=\pm \left(1+|g_{\mathbf{k}}^{{\lambda}\xi_s}|^2\right)^{-\frac{1}{2}} 
\end{equation}
with $g_{\mathbf{k}}^{{\lambda}\xi_s}=t^{\lambda_s} e(\mathbf{k})(\Delta\varepsilon^{\lambda}_{\xi_s}/2-\varepsilon^{\lambda}_{\mathbf{k},\xi_s})^{-1}.$
The appearing nearest-neighbor tight-binding function $e(\mathbf{k})=\sum_{j}^3 e^{i\mathbf{k}\cdot \mathbf{b}_j}$ can be Taylor-approximated for small momenta close to the K points resulting in $e(\mathbf{k})\approx\frac{\text{i}\sqrt{3}}{2}\text{k}e^{\pm\text i\phi_{\text{k}}}$ with the geometric phase $e^{\pm\text i\phi_{\text{k}}}$ (+/- standing for the K/K' point). This phase enters all coupling elements including the optical matrix element including optical selection rules and the Coulomb matrix element describing the formation of excitonic states. In combination, they give rise to the lifting of the degeneracy between 2s and 2p states and they determine, which excitonic states are optically bright (s-states) and dark (p-states).

The optical matrix element $M^{\sigma_\pm}_{\xi_s}(\mathbf{k})=\langle\Psi^{\lambda,\xi_s}(\mathbf k,\mathbf r)|p^{\pm}|
\Psi^{\lambda^\prime,\xi_s}(\mathbf k,\mathbf r)\rangle$ is the expectation value of the impulse $p^\pm$ projected to the direction of circularly polarized light. 
Inserting the tight-binding wave functions, we can obtain an analytic expression \cite{Bergh2014b} 
\begin{equation}
\label{eq:optmphase}
M^{\sigma_\pm}_{\xi_s}(\mathbf{k})=\bar{M}_{\text{k}}e^{\tau_{\xi} i\phi_\text{k}}(1\pm\tau_{\xi})
\end{equation}
with the geometric phase $e^{\tau_{\xi} i\phi_\text{k}}$, where $\tau_{\xi}=$+/-1 denotes the K/K' point. The strength of the coupling is determined by 
$\bar{M}_{\text{k}}=c_0C_{X,\mathbf{k}}^{v \xi_s *}C_{X,\mathbf{k}}^{c \xi_s}|\mathbf{k}|$ with $c_0=$ 1.04 meV that has been adjusted to obtain a light absorption of 10\% in agreement with recent measurements \cite{Moody2015}. The optical matrix element determines the linear response function $\chi(\omega)$ that is given by the macroscopic polarization for circularly left ($\sigma_-$) or right ($\sigma_+$) polarized light 
\begin{equation}
\begin{split}
 \label{macro}
P^{\sigma_{\pm}}(t)&=\sum_{n,m_l\mathbf k,\xi_s} \theta^{n,m_l}_{\xi_s \mathbf k}p^{n,m_l} (t)\,M^{\sigma_{\pm}*}_{\xi_{s}}(\mathbf{k})+cc\\\notag
&=\sum_{n,m_l,\text{k},\xi_s} p^{n,m_l} (t) \,\bar{M}_{\text{k}}
\sum_{\phi_{k}} 
\theta^{n,m_l}_{\xi_s \mathbf k}e^{-\tau_{\xi} i\phi_\text{k}}(1\pm\tau_{\xi})+cc.
\end{split}
\end{equation}
Besides the optical matrix element, the excitonic wave function $\theta^{n,m_l}_{\xi_s \mathbf k}$ and its angle dependence play a crucial role for the optical response. 
For isotropic systems, the appearing sum over the angle $\phi_k$ is only non-zero if its argument is symmetric. Since the azimuthal quantum number $m_l$ is representing the phase of the eigenfunction $e^{i m_l\phi_\text{k}}$, we can rephrase Eq. (\ref{macro}) by shifting $m_l$ by $\tau_\xi$ resulting in $m_l\rightarrow\bar{m_l}=m_l-\tau_\xi$ and $\theta^{n,m_l}_{\xi_s \mathbf k}e^{-\tau_{\xi} i\phi_\text{k}}=\theta^{n,m_l-\tau_{\xi}}_{\xi_s \mathbf k}=\theta^{n,\bar{m_l}}_{\xi_s \mathbf k}$.
One might argue that since 
we find bright states for $\bar{m_l}=0$ resulting in $m_l=\tau_{\xi}$, 
 the geometric phase leads to a non-zero optical response only for p-like excitonic states with $m_l\neq 0$. However, to obtain a complete picture, one has to take into account also the geometric phase in the Coulomb interaction.

In agreement with kp-theory and many-body Bethe-Salpeter approach \cite{Barry_PRL_Xiao2015,Barry_PRL_Imamoglu2015,Berkelbach2015}, the geometric phase enters the Coulomb matrix element $V_{\mathbf{k},\mathbf{k'},\mathbf{q}}^{\lambda\lambda'\xi_s}=\varGamma^{\lambda\lambda'\xi_s}_{\mathbf{k},\mathbf{k'},\mathbf{q}}V_{\mathbf{q}}$ via the overlap integrals appearing in the tight-binding coefficients 
$\varGamma^{\lambda\lambda'\xi_s}_{\mathbf{k},\mathbf{k'},\mathbf{q}}=\sum_{j,f=X,M}C_{f,\mathbf{k}}^{\lambda\xi_s *}C_{j,\mathbf{k'}}^{\lambda'\xi_s *}
C_{j,\mathbf{k'}+\mathbf{q}}^{\lambda'\xi_s}C_{f,\mathbf{k}-\mathbf{q}}^{\lambda\xi_s}$.
This term leads to an angular dependence of the Coulomb interaction and hence to the cancellation of the energy degeneration of excitonic states with the same excitonic index $n$, but different azimuthal quantum number $m_l$, cf. the supplementary material for more details.

To evaluate Eq. (\ref{macro}), we solve the Wannier equation to obtain the excitonic wave function $\theta^{n,m_l}_{\xi_s\mathbf k}$ and the Bloch equation to obtain the excitonic polarization $p^{n m_l}_{\xi_s\mathbf Q}$. The Wannier equation reads \cite{Bergh2014b}
\begin{equation}
 \tilde{\epsilon}_{\mathbf{k},\xi_s} \theta^{nm_l}_{\xi_s\mathbf k}
 -\sum_{\mathbf{k}^\prime}\varGamma^{vc\xi_s}_{\mathbf{k},\mathbf{k^\prime}}V_{\mathbf{k}-\mathbf{k^\prime}}\theta^{nm_l}_{\xi_s\mathbf{k}^\prime}=E^{n m_l}_{\xi_{s}}\theta^{nm_l}_{\xi_s\mathbf k}
 \label{eq:Wannier}
\end{equation}
with the excitonic eigenstates $\theta^{nm_l}_{\xi_s\mathbf k}$ and eigenvalues $E^{n m_l}_{\xi_{s}}$.
Here, the electronic dispersion relation $\tilde{\epsilon}_{\mathbf{k},\xi_s}=\epsilon^{c}_{\mathbf{k},\xi_s}-\epsilon^{v}_{\mathbf{k},\xi_s}$ appears as well as the tight-binding coefficient $\varGamma^{vc\xi_s}_{\mathbf{k} ,\mathbf{k^\prime}}$ that contains the geometric phase. To be more exact, the summation over atomic sublattice $j$ in $\varGamma^{vc\xi_s}_{\mathbf{k} ,\mathbf{k^\prime}}$ leads to three terms including one without any phase and two with phase expressed by $e^{\pm\tau_\xi i \phi_{\mathbf k-\mathbf k^\prime}}$ and weighted by differently strong Coulomb interactions within the the Keldish potential \cite{Keldysh1979}, cf. the supplementary material.

 Evaluating the Wannier equation for different azimuthal quantum numbers $m_l$,
 we find the corresponding excitonic eigenenergies and eigenfunctions, which are illustrated in Fig.\ref{fig:1}. 
To determine which states are optically active, we use the shifted $\bar{m_l}$ introduced above (containing the geometric phase from the optical matrix element) and consider the three different phase terms appearing in $\varGamma^{vc\xi_s}_{\mathbf{q},\mathbf{\bar{q}}}$. We find that one of these terms cancels the geometric phase from the optical matrix element, since
 $\theta^{n\bar{m_l}}_{\xi_s\mathbf k}e^{+\tau_{\xi} i\phi_\text{k}}=\theta^{n\bar{m_l}+\tau_{\xi}}_{\xi_s\mathbf k}=\theta^{nm_l}_{\xi_s\mathbf k}$ with $m_l=\bar{m_l}+\tau_{\xi}$, i.e. we have shifted the azimuthal quantum number back to normal order. This results in strongly bound and bright s excitonic states with $m_l=0$, cf. first row of Fig.\ref{fig:1} (a). 
 To put it in a nutshell, although the geometric phase lifts the degeneracy between excitonic states of different angular momentum, it does not change the optical selection rules.

 In agreement with recent experimental data \cite{Chernikov2014}, our calculations reveal that higher excitonic states are stronger bound than expected in the 2D Rydberg series. This can be traced back to the relatively weak Keldish screening of the Coulomb interaction in atomically thin TMDs \cite{Chernikov2014,Bergh2014b}. 
Furthermore, we find that due to the phase-dependent Coulomb interaction, the degeneracy between 2s and 2p states is lifted with 2p states now lying about 35 meV below the 2s states, which is also in line with recent studies \cite{Barry_PRL_Imamoglu2015,Barry_PRL_Xiao2015}.
The 2p states themselves are split by approximately 15 meV depending on the sign of the phase, cf. Fig. \ref{fig:1} (a). The latter is given by the azimuthal quantum number $m_l$ and prohibits their coupling to an external light field. This is manifested by their vanishing contribution to the macroscopic polarization in Eq. (\ref{macro}) due to the appearing integral over the angle $\phi_{\mathbf k}$. 

The energy difference of the 2p states is also reflected by a different radial component of the corresponding excitonic wave functions, cf. Figs. \ref{fig:1} (b) and (c).
Our calculations show that similarly to the ideal hydrogen model the stronger bound the excitonic state, the broader is its eigenfunction in the momentum space and the more localized it is in the real space.
 As a result, the 1s state is much broader in momentum space compared to the 2s state, while the stronger bound 2p$_-$ exciton is slightly broader than the 2p$_+$ exciton, cf. Figs. \ref{fig:1} (b) and (c).
 
Having determined the excitonic eigenvalues and eigenfunctions, we have all ingredients to formulate TMD Bloch equations for the time- and momentum-dependent microscopic polarization within the excitonic basis. 
Since the disorder-exciton coupling induces polarization $p^{vc}_{\xi_s\mathbf k_i \mathbf k_j }$ with $\mathbf k_j\neq \mathbf k_i$, we introduce the relative momentum $\mathbf{q}=\alpha \mathbf k_{i}+\beta \mathbf k_{j}$ and the center-of-mass momentum $\mathbf{Q}=\mathbf k_{i}- \mathbf k_{j}$ with $\alpha=\frac{m_{\lambda_i}}{m_{\lambda_i}+m_{\lambda_j}}$ and $\beta=\frac{m_{\lambda_j}}{m_{\lambda_j}+m_{\lambda_i}}$, where $m_{\lambda_i, \lambda_j}$ is the effective mass of the band $\lambda_i, \lambda_j$, taken from Ref. \onlinecite{Ramasubramaniam2012}. Then, the transformation of the microscopic polarization 
into the excitonic basis 
is more general yielding
$p^{vc}_{\xi_s\mathbf{q},\mathbf{Q}}(t)=\sum_{n,m_l}p^{nm_l}_{\xi_s\mathbf Q} \theta^{nm_l}_{\xi_s\mathbf q}$. The corresponding TMD Bloch equation reads:
\begin{align}
\label{eq:p_excitonic}
i\hbar \dot{p}^{n m_l}_{\xi_s\mathbf Q} (t)&=
\left(E_{\xi_s\mathbf Q}+E^{n m_l}
_{\xi_s}+i\gamma_{\mathbf Q}^{\text{hom}}\right) p^{n m_l }_{\xi_s\mathbf Q}(t)
\\\nonumber
&+\delta_{\mathbf Q,0}\delta_{m_l,0}\Omega^{n\pm}(t)
+\sum_{\mathbf Q^\prime \nu \mu_l} D_{\xi_s\mathbf Q^\prime}^{\nu \mu_l} p^{\nu \mu_l}_{\xi_s\mathbf Q-\mathbf Q^\prime}(t).
\end{align}
Here, the excitonic dispersion is given by $E_{\xi_s\mathbf Q}=\hbar\mathbf Q^2/(2M)$ with $M=m_e+m_h$. Since the focus of our work lies on a qualitative investigating of fundamental aspect of exciton-disorder coupling, we do not include the effects of the Coulomb exchange coupling on excitonic dispersion \cite{Barry_PRL_Louie2015, MacDonald2015,Wang2014}. 
The TMD Bloch equation allows us to calculate the excitonic absorption as well as to study the impact of exciton-disorder coupling on optical selection rules. It includes all excitonic states with eigenenergies $E^{nm_l}_{\xi_s}$, the coupling of the system with incoming light determined by the Rabi frequency 
$\Omega^{n\pm}(t)=\sum_{\mathbf q}A(t)M^{\sigma_{\pm}}(\mathbf{q})\theta^{nm_n=0 *}_{\xi_s\mathbf q}$, and exciton-disorder coupling determined by the matrix element $D_{\xi_s\mathbf Q^\prime}^{\nu \mu_l}$. 
To account for homogeneous broadening at room temperature via higher-order phonon-induced scattering processes, we introduce a phenomenological dephasing $\gamma_{\mathbf Q}^{\text{hom}}$, which has been set to 12 meV for all $\mathbf Q$ in agreement with our recent calculations \cite{Malte2016a} and experimental observations \cite{Moody2015}.

Due to the small momentum of photons, we consider no momentum transfer via optical excitation ($\delta_{\mathbf Q,0}$ in Eq. (\ref{eq:p_excitonic})) and hence only symmetric excitonic states ($m_l=0$) without center of mass momentum ($\mathbf Q=0$) couple to an external light field. However, the exciton-disorder term in Eq. (\ref{eq:p_excitonic}) couples p- and s-like excitonic states, which crucially changes the optical selection rules. Since in TMDs the 2s and 2p states are not degenerate due to the non-vanishing geometric phase, the disorder-induced coupling is expected to lead to pronounced effects even in linear absorption spectra.
The disorder contribution is averaged on the macroscopic level by calculating the microscopic polarization for a large number of random disorder realizations. The method has been implemented following the approach of Glutsch and Bechstedt \cite{BechstedtPRB1994}. The exciton-disorder matrix element entering Eq. (\ref{eq:p_excitonic}) reads 
\begin{equation}
\begin{split}
\nonumber
 D_{\xi_s\mathbf Q^\prime}^{\nu \mu_l}&= \sum_{\mathbf q \lambda}U(\mathbf Q^\prime) \Gamma^{\lambda\lambda}_{\xi_s\mathbf q,\mathbf Q,\mathbf Q^\prime}\theta_{\xi_s\mathbf q}^{n m_l *}
 \left(\theta_{\xi_s \mathbf q+\beta \mathbf Q^\prime}^{\nu\mu_l} +\theta_{\xi_s \mathbf q-\alpha \mathbf Q^\prime}^{\nu\mu_l}\right)
\end{split}
\end{equation}
with $\mathbf q$ describing the relative and $\mathbf Q, \mathbf Q^\prime$ the center-of-mass motion. 
The disorder potential $U(\mathbf Q^\prime)=R(\mathbf Q^\prime)Ae^{\frac{1}{4}\lambda_{c}^{2} Q^{\prime 2}}$ is determined by the amplitude $A$ giving the maximal strength of the disorder, the correlation length $\lambda_c=1$ nm according to the excitonic Bohr radius \cite{RanaLT2016}, and $R(\mathbf Q^\prime)$ being a random number on the complex unitary circle fulfilling the condition $R(\mathbf Q^\prime)=R(-\mathbf Q^\prime)^*$ \cite{BechstedtPRB1994}.
Furthermore, the tight-binding coefficients directly enter in 
$\Gamma^{\lambda\lambda}_{\xi_s\mathbf q,\mathbf Q,\mathbf Q^\prime}=\sum_{j=X,M}C^{\lambda\xi_s *}_{j}(\mathbf q+a_\lambda \mathbf Q) C^{\lambda\xi_s}_{j}(\mathbf q+\mathbf Q^\prime+a_\lambda \mathbf Q)$ 
with $a_v=\beta$ and $a_c=-\alpha$.
For perfect crystals or for isolated atoms, one finds 
$\sum_{\mathbf q}\theta_{\xi_s\mathbf q}^{n m_l}\theta_{\xi_s\mathbf q}^{\nu \mu_l *}=\delta_{n,\nu}\delta_{m_l,\mu_l}$ 
due to the orthogonality of the wave functions. The term vanishes for wave functions with $m_l\neq \mu_l$, since the phase difference results in $\sum_{\phi}e^{i (m_l\phi_{\mathbf q-\mathbf Q^\prime}-\mu_l\phi_{\mathbf q})}=0$. 
However, under the influence of disorder the microscopic quantities gain a center-of-mass momentum, which enters the wave function and breaks the symmetry of the angular integral. It is well known that in a systems with center-of-mass motion the azimuthal quantum number is not a conserved quantity \cite{Zhu2014}.
Here, the sum over the angular momentum is not symmetric resulting in $\sum_{\phi}e^{i(m_l\phi_{\mathbf q-\mathbf Q^\prime}-\mu_l\phi_{\mathbf q})}\neq0$. Therefore, the exciton-disorder matrix element does not vanish and leads to a coupling between optically bright s- and dark p-like excitonic states.
In general, one can state that center-of-mass motion destroys the angular symmetry leading to a non-vanishing coupling between states of different angular momentum. 
\begin{figure}[t!]
\includegraphics[width=1.\linewidth]{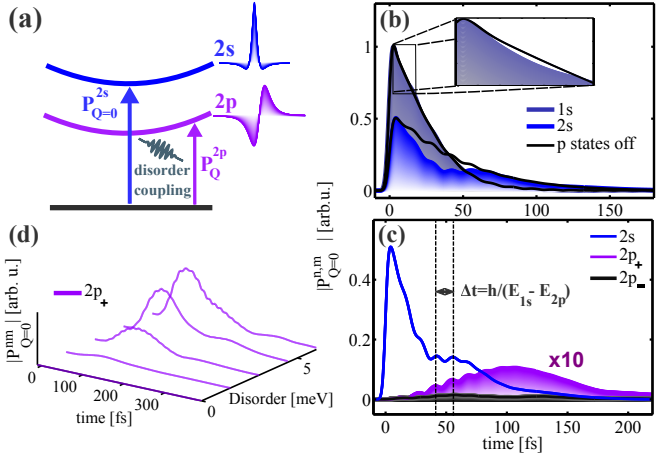}
 \caption{Microscopic polarizations of higher excitonic states. (a) Schematic illustration of the optically excited 2s polarization $p^{2s}_{\mathbf Q=0}$ and the generation of disorder-induced 2p polarization $p^{2p}_{\mathbf Q}$. (b)  Temporal evolution of the absolute value of  $p^{1s}_{\mathbf{Q}=0}$ and $p^{2s}_{\mathbf{Q}=0}$, where the black lines show the case without the coupling to 2p states. (c)
Direct comparison between the polarization of 2s, 2p$_+$, and 2p$_-$ states.
 (d) Temporal evolution of $p^{2p_+}_{\mathbf Q=0}$ as a function of disorder strength.
}
\label{fig:2}
\end{figure}

Now, we numerically solve Eq. (\ref{eq:p_excitonic}) for a large number of disorder realizations until we obtain a converged result. 
Figure \ref{fig:2} shows the temporal evolution of the disorder-averaged microscopic polarizations $p^{1s}_{\mathbf Q=0}$, $p^{2s}_{\mathbf Q=0}$, and $p^{2p_\pm}_{\mathbf{Q}=0}$ 
after applying a weak, spectrally broad pulse excitation.
The microscopic polarizations of the s states are optically driven via the external vector potential $A(t)$. 
These bright states induce the initially dark p-type polarizations $p^{2p_\pm}$ via exciton-disorder coupling, cf. Fig. \ref{fig:2} (a) and Eq. (\ref{eq:p_excitonic}).
The optically driven and disorder-induced polarizations couple to each other resulting in an oscillation transfer between the states. 
To illustrate this, Figs. \ref{fig:2} (b)-(d) show the absolute value of  the microscopic polarizations. The frequency of the relatively fast oscillation of 2p$_\pm$ polarizations  in Fig. \ref{fig:2} (c) corresponds to $\omega=2\pi(E_{1s}-E_{2p_\pm})/\hbar$ 
reflecting the central role of the optically driven 1s polarization. Calculating $p^{1s/2s}_{\mathbf Q=0}$ without exciton-disorder coupling (black lines in Fig. \ref{fig:2} (b)) illustrates a slower dephasing of the 2s polarization.

The crucial quantity for the occurrence of these effects is the disorder strength. Depending on the quality of the material, the degree of disorder (reflected by the  inhomogeneous broadening)  can strongly vary. Most samples investigated so far show inhomogeneous line widths at low temperatures of some tens of meV \cite{Moody2015}.
In line with these investigations, we assume a disorder strength of 6.5 meV. Here, the disorder couples states with center-of-mass momentum to the optically excited state resulting in multiple resonances and an an inhomogeneous broadening of the 1s excitonic resonance of 30 meV.
To investigate the impact of disorder, Fig.\ref{fig:2} (d) shows the temporal evolution of $p^{2p_+}$ as a function of disorder. The stronger the disorder, the more pronounced is the induced microscopic polarization of the initially dark 2p state and the weaker is the decay of the appearing oscillations.

\begin{figure}[t!]
 \begin{center}
\includegraphics[width=1.\linewidth]{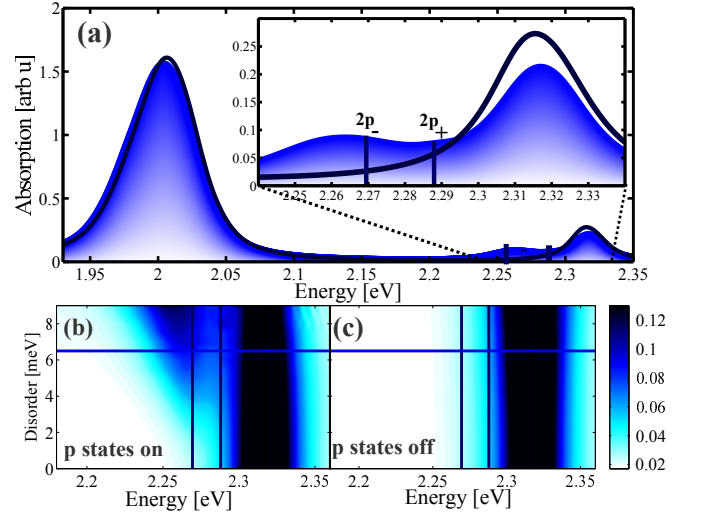}
 \end{center}
 \caption{Absorption spectrum of WS$_2$. (a) The blue-shaded spectrum shows the absorption including disorder-induced bright-dark-state coupling. 
For comparison, the black line describes the spectrum without the coupling demonstrating a clearly enhanced absorption in the vicinity of the 2p states (cf. the inset) and a red-shift of 1s and 2s resonances. 
Contour plot (b) with and (c) without exciton-disorder coupling showing the absorption  as a function of energy and disorder strength illustrating the increasing absorption 
in the vicinity of the 2p states (vertical lines) for higher disorder. The horizontal blue line marks the case plotted in 
part (a).  
}
\label{fig:3}
\end{figure}
Next, we study the influence of the disorder-induced polarizations of dark 2p states on the experimentally accessible optical absorption of the exemplary WS$_2$ material. 
Figure \ref{fig:3} illustrates the spectrum in the vicinity of the 1s and 2s (inset) excitonic resonance in the presence of disorder.
Although p-like polarizations with $m_l\neq0$ do not directly contribute to the absorption spectrum 
(cf. Eq. (\ref{macro})) they do leave visible fingerprints via exciton-disorder coupling. We observe a clearly enhanced absorption at the position of 2p$_\pm$ excitonic states, cf. the inset of Fig. \ref{fig:3}. How pronounced this effect is, depends on the disorder strength, as illustrated in the contour plot in Fig.\ref{fig:3}(b). The stronger the disorder, the larger is $p^{2p_\pm}_{\mathbf Q}$ (cf. Fig. \ref{fig:2}(d)), and the more pronounced is the enhancement of the absorption appearing at the position of 2p$_\pm$ excitons and the larger is the red-shift of the 2s excitonic peak.

The appearance of additional oscillator strength can be understood in a straightforward way by reducing the complexity of the equations to the most important aspects: Considering a situation of two states $E_1, E_2$ coupled via $C_{12}$, where only the energetically lower state $E_1$ is driven by an external source, we find an analytic solution for resonances appearing in the absorption spectrum
$
 \hbar\omega_\pm=(E_1+E_2)/2\pm\sqrt{\left(\left(E_{1}-E_{2}\right)/2\right)^2+|C_{12}|^2}.
$ 
Although the state $E_2$ is not directly driven, we find a resonance at $\hbar\omega_{+}=E_2$ for $(E_{1}-E_{2})^2\gg |C_{12}|^2$. 
 This is exactly what happens when exciton-disorder interaction couples the optically active 1s state to the initially dark 2p states.
Note that the additional resonances found in the spectrum can be expected to be more pronounced at low temperatures where the homogeneous broadening of the main 2s peak via phonons is smaller allowing a better resolution of the weak disorder-induced peaks. 

In conclusion, we have presented a microscopic study on the impact of exciton-disorder coupling on the temporal dynamics and the absorption spectra of transition metal dichalcogenides. We find that the non-vanishing geometric phase lifts the degeneracy between 2s and 2p states and that the significant exciton-disorder interaction softens optical selection rules in these materials. It induces a microscopic polarization of optically dark 2p$_\pm$ states leaving an optical fingerprint in linear absorption spectra. The gained insights on the impact of exciton-disorder coupling on optical selection rules are of general nature and therefore applicable to related materials with a strong Coulomb interaction and significant degree of disorder.
\begin{acknowledgments}
We acknowledge financial support from the EU Graphene Flagship (no. 604391) and the Deutsche Forschungsgemeinschaft (DFG) through SPP 1459  and SFB 951.
\end{acknowledgments}

\end{document}